\documentstyle[aps,prb,multicol,psfig]{revtex} \newif\ifmynarrow
\mynarrowfalse
\renewcommand{\narrowtext}{%
  \ifmynarrow\hspace*{\fill}\raisebox{-1ex}[0pt][0pt]{%
    \rule{0.3pt}{1ex}%
    \rule[1ex]{20.5pc}{0.3pt}}\fi \mynarrowtrue
  \vspace{-1.0ex}%
  \begin{multicols}{2}%
  \par\global\columnwidth20.5pc
  \global\hsize\columnwidth\global\linewidth\columnwidth
  \global\displaywidth\columnwidth}
\renewcommand{\widetext}{%
  \end{multicols}%
  \vspace{-2.5ex}%
  \noindent\raisebox{1ex}[0pt][0pt]{%
    \rule{20.5pc}{0.3pt}%
    \rule{0.3pt}{1ex}}%
  \par\global\columnwidth42.5pc
  \global\hsize\columnwidth\global\linewidth\columnwidth
  \global\displaywidth\columnwidth}
\begin{document}
\draft \tighten

\title{Longitudinal spin decoherence in spin diffusion in
  semiconductors} 
\author{M. Q. Weng$^b$ and M. W. Wu$^{a,b,\ast}$}
\address{$^{a}$Structure Research Laboratory, 
  University of Science \& Technology of China, Academia Sinica,  Hefei, Anhui, 230026,
  China}
\address{$^{b}$Department of Physics,
  University of Science \& Technology of China, Hefei, Anhui, 230026,
  China$^{\ast\ast}$}
\date{\today}
\maketitle
\begin{abstract}
We  set  up a set of many-body kinetic Bloch equations
with spacial inhomogeneity. We reexamine  the widely adopted
quasi-independent electron model and  show the inadequacy of this model in
studying the spin transport.  We further  point  out a new decoherence effect based 
on  interference along the direction of diffusion in spin
  transport due to the so called inhomogeneous broadening effect in the Bloch
equations. We  show  that this inhomogeneous  broadening can cause 
decoherence alone, even in the absence of  the scattering and that
the resulting decoherence is more important than the dephasing effect
due to both the D'yakonov-Perel' (DP) term and the scattering.
\end{abstract}

\pacs{PACS: 72.25.Dc; 72.25.Rb}

\narrowtext 
Study of spintronics has attracted tremendous attention in
recent years, both in theoretical and experimental circles,\cite{wolf}
thanks to the discovery of the long-lived (sometimes $>$ 100\ ns)
coherent electron spin states in $n$-typed
semiconductors.\cite{kikk1,kikk2,kikk3,hohno,ohno1,kikk4} Possible
applications of spintronics include qubits for quantum computers,
quantum memory devices, spin transistors, and spin valves etc. The
last two applications involve transporting 
spin polarized electrons  from a place to another by means of an electrical or
diffusive current. Therefore, it is of great importance to study the
spin transport.  Apart from the great number of works on spin
injection, there are only a few experimental reports on 
coherent spin transport over macroscopic
distance.\cite{kikk2,malajovich_prl_2000,jonker_prb_2000}
On theoretical aspect, most works are based on  a quasi-independent
electron model and focused on the diffusive transport
regime,\cite{schmidt,spintronics,flatte_prl_2000,zutic_0106085,%
Yuzg_0201425,martin_0201481} where equations for spin polarized
currents can be set up and the longitudinal spin dephasing, generally
referred to as spin diffusion length can be achieved.  In these
theories, the mechanism for the spin relaxation is assumed  to be
due to the  spin-flip scattering.  In the absence of  the scattering, the spin
polarization will not decay in  a  nonmagnetic sample.  In
Ref. \onlinecite{takahashi_prb_1999}, Takahashi {\it et al.}
calculated the scattering induced spin relaxation time associated with
the spin diffusion starting from the many body kinetic equations.

Of particular interest to the spin transport theory in semiconductors
has been the question as to whether the quasi-independent electron model
can adequately account for the experimental results or whether
many-body processes are important.
Flatte {\em et al.} have concluded
that an independent electron approach is quite capable of explaining
measurements of spin lifetimes in the diffusive regime.\cite{flat} In
this paper, we reexamine this issue from a full many-body transport
theory and show the inadequacy of the independent electron model in
describing the spin transport.  We also propose a mechanism that may
cause strong longitudinal spin decoherence in addition to the spin
dephasing due to  scattering. The new mechanism
is based on  the interference effect due to the 
wavevector dependence of the spin densities along the spacial gradients 
in the spin diffusion.  This wavevector dependence can be considered as some sort of
``inhomogeneous broadening'', which can cause spin decay alone, even
in the absence of scattering.

Recently, we have presented a many-body kinetic theory to
describe the spin precession and dephasing in insulating  samples  as well
as $n$-doped samples.\cite{wu1,wu2,wu3} In this paper we extend this
theory to the spacial inhomogeneous regime and  obtain  the many-body
transport equations  necessary to investigate the spin diffusion in n-doped GaAs.
 Here, we only focus on the spin transport inside the 
semiconductor  and avoid the problem of spin injection at the
boundary. Based on the 
two-spin-band model\cite{wu2} in the conduction bands, we construct
the semiconductor Bloch equations by using the nonequilibrium Green
function method with gradient expansion as well as  the generalized
Kadanoff-Baym Ansatz\cite{haug} as follows: 
\widetext
\begin{equation}
\label{eq1}
{\partial\rho({\bf R},{\bf k},t)\over\partial t} -{1\over 2}
\bigl\{{\bf \nabla}_{\bf R}\bar{\varepsilon}({\bf R},{\bf k}, t), {\bf
\nabla}_{\bf k}\rho({\bf R},{\bf k},t)\bigr\} +{1\over 2} \bigl\{{\bf
\nabla}_{\bf k}\bar{\varepsilon}({\bf R},{\bf k},t), {\bf \nabla}_{\bf
R}\rho({\bf R},{\bf k},t)\bigr\} -\left.{\partial\rho({\bf R},{\bf
k},t)\over\partial t}\right|_{c}\nonumber\\ =\left.{\partial\rho({\bf
R},{\bf k},t)\over\partial t}\right|_{s}.
\end{equation}
\narrowtext
\noindent Here $\rho({\bf R},{\bf k}, t)$ represents a
single particle density matrix. The diagonal elements describe the
electron distribution functions $\rho_{\sigma\sigma}({\bf R},{\bf
k},t)=f_{\sigma}({\bf R},{\bf k},t)$ of wave vector ${\bf k}$ and spin
$\sigma$($=\pm 1/2$) at position ${\bf R}$ and time $t$. The
off-diagonal elements $\rho_{\sigma-\sigma}({\bf R},{\bf k}, t)$
describe the inter-spin-band polarization components (coherences) for
the spin coherence. The quasi-particle energy
$\bar{\varepsilon}_{\sigma\sigma^{\prime}}({\bf R},{\bf k},t)$, in the
presence of a moderate magnetic field ${\bf B}$ and with the DP
mechanism\cite{dp} included, can be written as
\begin{eqnarray}
  \label{eq:energy}
\bar{\varepsilon}_{\sigma\sigma^{\prime}}({\bf R},{\bf k},t)&=&
  \varepsilon_{k}\delta_{\sigma\sigma^{\prime}} +\bigl[g\mu_B{\bf
  B}+{\bf h}({\bf k})\bigr]\cdot {\vec{\bf
  \sigma}_{\sigma\sigma^{\prime}}\over 2}\nonumber\\ &&\mbox{}-
  e\psi({\bf R},t) +\Sigma_{\sigma\sigma^{\prime}}({\bf R},{\bf
  k},t)\;.
\end{eqnarray}
Here $\varepsilon_k=k^2/2m^\ast$ is the energy spectrum with $m^\ast$
denoting electron effective mass,   $-e$ is the electron charge
  and  $\vec{\bf \sigma}$ are the Pauli 
matrices and ${\bf h}({\bf k})$ originate from the DP mechanism which
contains both the Dresselhaus\cite{Dress} and the Rashba
terms.\cite{ras} In this paper, we only consider the first one.  For
[001] quantum well, it can be written as\cite{epp}
$h_x({\bf k})=\gamma k_x(k_y^2-\kappa_z^2)$, $h_y({\bf k})=\gamma
k_y(\kappa_z^2-k_x^2)$,
with $\kappa_z^2$ denoting the average of the operator
$-(\partial/\partial z)^2$ over the electronic state of the lowest
subband. $\gamma=(4/3)(m^\ast/m_{cv})(1/\sqrt{2m^{\ast
3}E_g})(\eta/\sqrt{1-\eta/3})$ and $\eta=\Delta/(E_g+\Delta)$. Here
$E_g$ denotes the band gap, $\Delta$ represents the spin-orbit
splitting of the valence band, and $m_{cv}$ is a constant close in
magnitude to  the free electron mass $m_0$.\cite{aro} The electric
potential $\psi({\bf R},t)$ satisfies  the Poisson equation
\begin{equation}
\label{poission}
{\bf \nabla}_{\bf R}^2\psi({\bf R},t)= -{e\bigl[n({\bf R},t)-n_0({\bf
R})\bigr]/\epsilon}\ ,
\end{equation}
where $n({\bf R},t)=\sum_{\sigma{\bf k}}f_{\sigma}({\bf R},{\bf k},t)$
is the electron density at position ${\bf R}$ and time $t$, and 
$n_0({\bf R})$ is the background positive electric charge density.
$\Sigma_{\sigma\sigma^{\prime}} ({\bf R}, {\bf k}, t) =-\sum_{{\bf
q}}V_{{\bf q}} \rho_{\sigma\sigma^{\prime}}({\bf R},{\bf k}-{\bf
q},t)$ is the Hartree-Fock self-energy, with $V_{\bf q}$ denoting the
Coulomb matrix element.  In 2D case, $V_{\bf q}$ is given by
\begin{equation}
  \label{eq:vq}
  V_{\bf q} = \sum_{q_z}{4\pi e^2\over\epsilon_0({\bf
  q}^2+q_z^2+\kappa^2)}|I(iq_z)|^2,
\end{equation}
in which $\kappa=2e^2m^{\ast}/\epsilon_0\sum_{\sigma}f_{\sigma}(K=0)$
is the inverse screening length,  with $\varepsilon_0$ being  the static
dielectric constant.  The form factor 
$|I(iq_z)|^2=\pi^2\sin^2y/[y^2(y^2-\pi^2)^2]$ with $y=q_za/2$.  
It is noted that when one takes only the diagonal elements
$\rho_{\sigma\sigma}$ of Eq.\ (\ref{eq1}) and neglects all
off-diagonal ones $\rho_{\sigma-\sigma}$, the first three terms on the
left hand side of the equation correspond to the drift terms in the
classical Boltzmann equation, modified with the DP terms and self
energy from  the  Coulomb Hartree term. 
${\partial \rho({\bf R},{\bf k},t) \over \partial t}|_c$ 
and 
${\partial \rho({\bf R},{\bf k},t) \over \partial t}|_s$ 
in the Bloch equations (\ref{eq1}) are the coherent and scattering
terms respectively, with the symbols $|_c$ and $|_s$ standing for
``coherent'' and ``scattering''. The components of the
coherent terms can be written as\cite{haug,wu3}
\begin{eqnarray}
  \left.{\partial f_{\sigma}\over \partial t}\right|_c&=&
  -2\mbox{Im}\bigl[\bar{\varepsilon}_{\sigma-\sigma}\rho_{-\sigma\sigma}\bigr],
  \label{eq:fc}\\
  \left.{\partial \rho_{\sigma-\sigma}\over \partial t}\right|_c&=&
  i\bigl[\bar{\varepsilon}_{\sigma\sigma}
  -\bar{\varepsilon}_{-\sigma-\sigma}\bigr] \rho_{\sigma-\sigma}
  +i\bar{\varepsilon}_{\sigma-\sigma} \bigl[f_{-\sigma}-f_{\sigma}
  \bigr]\ .
  \label{eq:rhoc}
\end{eqnarray}
While the scattering terms $\left.{\partial\rho({\bf R},{\bf
k},t)\over\partial t}\right|_{s}$ are given in detail in Eqs.\ (5) and
(7) of Ref. \onlinecite{wu3}.

The Bloch equations (\ref{eq1}) can be reduced to 
their counterpart  in the independent electron approach as follows. 
The DP term forms an effective magnetic field. It can flip
the spin-up electrons to the spin-down ones, and 
{\em  vice versa}. 
The DP term combines with the scattering will result in a
longitudinal spin dephasing.\cite{dp,wu2,wu3} By applying the relaxation time
approximation to describe this dephasing and discarding the spin
coherences $\rho_{\sigma-\sigma}({\bf R},{\bf k},t)$ as well as the DP
term (to avoid double counting) and carrying out the
summation over ${\bf k}$, one obtains the
the continuity equation for electrons of spin $\sigma$
\begin{equation}
   \label{eq:continuity}
{\partial n_{\sigma}({\bf R},t)\over \partial t}- {1\over e}{\bf
\nabla}_{\bf R}{\bf \cdot}{\bf J}_{\sigma}({\bf R},t)=
-{n_{\sigma}({\bf R},t)-n_0({\bf R},t)\over \tau_s}\ ,
\end{equation}
in which  $n_0({\bf
  R},t)=[n_{\sigma}({\bf R},t)+n_{-\sigma}({\bf R},t)]/2$
is the total electron number at ${\bf R}$.    ${\bf J}_{\sigma}({\bf R},t)=\sum_{{\bf k}}(-e){\bf
  v}_{\sigma{\bf k}}f_{\sigma}({\bf R},{\bf k},t)$ is the electric
current of spin $\sigma$. 
The spin dependent velocity is ${\bf
  v}_{\sigma{\bf k}}={\bf \nabla}_{\bf
  k}\bar{\varepsilon}_{\sigma\sigma}({\bf R}, {\bf k}, t)$ where
$\bar{\varepsilon}_{\sigma\sigma}({\bf R}, {\bf k}, t)$ is given by
Eq.\ (\ref{eq:energy}) but without the DP term ${\bf h}({\bf k})$.
By applying the relaxation time approximation to
describe the momentum scattering and keeping terms of the 
the lowest order ({\em ie.}, neglecting terms containing
$\rho_{\sigma-\sigma}$) and carrying out the summation over ${\bf k}$, 
one obtains the expression for the current in the steady state:
\begin{equation}
  \label{eq:current}
  {\bf J}_{\sigma}({\bf R}, t)=n_{\sigma}({\bf R},t)e\mu{\bf E}({\bf
    R},t) + eD\nabla_{\bf R}n_{\sigma}({\bf R},t)\ .
\end{equation}
Here  $\mu$ and $D$ represent the electron mobility and
diffusion  constant respectively.  Equations (\ref{eq:continuity}) and
(\ref{eq:current}) are the diffusion equations in the independent
electron approach.\cite{schmidt,spintronics,zutic_0106085,Yuzg_0201425,martin_0201481}

One can see from the derivation of  above diffusion equations
that, by summing over ${\bf k}$, the ${\bf k}$ dependence of the coefficients
of ${\bf \nabla}_{\bf R}\rho({\bf R},{\bf k},t)$ in the Bloch equation
(\ref{eq1}) is removed.  This will not cause any problem when there is
no  spin precession.   However, when the electron spin 
precesses along with  the diffusion, {\em e.g.}
 in the presence of a magnetic field or  of
an effective one ({\em i.e.} the DP term), this kind of ${\bf k}$
dependence may cause additional decoherence.

To reveal this effect , we apply the above kinetic equation to study
the stationary state in the plane of  an $n$-doped GaAs quantum
well (QW), with its growth direction along the $z$-axis. The width of the QW is
assumed to be small enough so that only the lowest subband is
important.  We assume one side of the sample ($x=0$) is connected with
an Ohmic contact which gives constant spin polarized injection.  In
this study, we assume the electric field $E=0$. The diffusion is along
the $x$ direction.  The electron distribution functions at the
interface are assumed to be  the Fermi distributions
\begin{equation}
  \label{eq:fk_0}
  f_{\sigma}(0,{\bf k},t)\equiv f^0_{\sigma}({\bf k})=
\{\exp[(\varepsilon_k-\mu_\sigma)/T]+1\}^{-1}\;,
\end{equation}
with $T$ being the temperature and $\mu_\sigma$ representing the
electron chemical  potential of spin $\sigma$.  The spin coherence at the interface is
assumed to be zero
\begin{equation}
  \label{eq:rho_0}
  \rho_{\sigma-\sigma}(0, {\bf k},t)\equiv 0\;.
\end{equation}
It is understood that the boundary condition here is an approximation
to describe the distributions just after  the injection of the {\em
  spin polarization}  from the Ohmic contact. There is no net charge
injection into the QW and the well is kept charge 
 neutral  everywhere.    
Actually, this boundary condition does not necessarily come from the
injection at the interface. It can also be produced in the center of
semiconductors by a circularly polarized cw laser. 

We first consider a much simplified case by neglecting the DP terms
${\bf h}({\bf k})$, the self energies as well as the scattering terms
in the Bloch equations (\ref{eq1}).  The simplified equations are
therefore as follows
\begin{eqnarray}
  \label{eq:f_sigma}
&&{k_x\over m^\ast}\partial_x f_{\sigma}(x,{\bf k})-
  g\mu_BB\mbox{Im}\bigl[\rho_{-\sigma,\sigma}(x,{\bf k})\bigr] = 0,\\
  \label{eq:rho_sigma}
&&{k_x\over m^\ast}\partial_x \rho_{\sigma-\sigma}(x,{\bf k})-
  i{g\mu_BB\over 2}\Delta f_{\sigma}(x,{\bf k}) = 0.
\end{eqnarray}
Here we take the magnetic field ${\bf B}$ along the $x$-axis. $\Delta
f_{\sigma}(x,{\bf k}) =f_{\sigma}(x,{\bf k})-f_{-\sigma}(x,{\bf k})$.
The solution for this simplified equations with the boundary
conditions (\ref{eq:fk_0}) and (\ref{eq:rho_0}) can be written out
directly
\begin{eqnarray}
  \label{eq:fs}
&& \Delta f_{\sigma}(x,{\bf k})=\Delta f^0({\bf k})\cos{g\mu_BBm^\ast
  x\over k_x}\;,\\
  \label{eq:rhos}
&&\rho_{\sigma-\sigma}(x,{\bf k})={i\over 2}\Delta f^0({\bf
  k})\sin{g\mu_BBm^\ast x\over k_x}\;.
\end{eqnarray}
Equations (\ref{eq:fs}) and (\ref{eq:rhos}) clearly show the effect of
the $k$-dependence to the spin precession along the diffusion
direction. For each fixed $k_x$, the spin precesses along the
diffusion direction with fixed period without any decay. Nevertheless,
for different $k_x$ the period is different. The total difference of
the electron densities with different spin is the summation  over
 all  wavenumbers $\Delta N=\sum_{\bf k}\Delta f_{\sigma}(x,{\bf k})$. It is
noted that the phase at the contact $x=0$ for different $k_x$ is all
the same. However, the speed of the phase of spin precession is
different for different $k_x$.  Consequently, when $x$ is large
enough, spins with different phases may cancel each other.  This can
further be seen from Fig.\ 1 where the electron densities $N_\sigma=
\sum_{\bf k}f_{\sigma}(x,{\bf k})$ for up and down spin are plotted as
functions of position $x$. The boundary electron densities at $x=0$
are $N_{1/2}(0)=2.05\times 10^{11}$\ cm$^{-2}$ and
$N_{-1/2}(0)=1.95\times 10^{11}$\ cm$^{-2}$. 
 We take $B=1$\ T and $T=200$\ K.  In order to
show the transverse spin dephasing, we plot in the same figure
the incoherently summed spin coherence $\rho(t)=\sum_{\bf
k}|\rho_{\frac{1}{2}-\frac{1}{2}}(x,{\bf k})|$.  
It is understood that
both  the true dissipation and the interference among the ${\bf
  k}$ states  may contribute to the decay. 
The decay due to interference is caused by 
 the different precessing rates of electrons with different
 wavevectors. 
For finite
system, this leads to reversible loss of coherence among
electrons.\cite{rauch,huxuedong} 
We refer to this kind 
of loss of coherence as decoherence. 
Whereas for the true  dissipation, 
the coherence of the
electrons is lost irreversibly.\cite{kuhn,rauch,huxuedong}
The irreversible loss of coherence is termed dephasing in this
paper. The incoherent summation is therefore
used to isolate the irreversible decay from the decay caused by
interference.\cite{wu2,kuhn} From the figure, one can see clearly the
longitudinal decoherence caused by the interference effect. It is also
noted from the figure that $\rho$ does not decay with the distance.
This is consistent with the fact that there is no  scattering in Eqs.\ 
(\ref{eq:fs}) and (\ref{eq:rhos}) and the decay comes  only from
the interference effect.

Facilitated with  the above understanding, we turn to the spin diffusion
problem with the DP terms, self-energies and  scattering 
included.  We take ${\bf B}={\bf E}=0$.  
By substituting the quasi-particle energy
$\varepsilon_{\sigma\sigma^{\prime}}({\bf R},{\bf k},t)$
[Eq.\ (\ref{eq:energy})] into the Bloch equations
(\ref{eq1}), the first three terms in Eqs.\ (\ref{eq1}) can
be written as
\widetext
\begin{eqnarray}
  \label{eq:dirver}
&&  \partial_t \rho_{\sigma\sigma^{\prime}}({\bf R},{\bf k},t)
  +e\partial_x \psi({\bf R},t)\partial_{k_x}
  \rho_{\sigma\sigma^{\prime}}({\bf R},{\bf k},t)
  -{1\over 2}\sum_{\sigma_1}\Bigl[
  \partial_x\Sigma_{\sigma\sigma_1}({\bf R},{\bf k},t)
  \partial_{k_x}\rho_{\sigma_1\sigma^{\prime}}({\bf R},{\bf k},t)
  +\partial_{k_x}\rho_{\sigma\sigma_1}({\bf R},{\bf k},t)
 \partial_x\Sigma_{\sigma_1\sigma^{\prime}}
({\bf R},{\bf k},t)\Bigr]\nonumber\\
&&+{k_x \over m}\partial_x
\rho_{\sigma\sigma^{\prime}}({\bf R},{\bf k},t)
+{1\over 4}\Bigl[\partial_x\bigl(h_x({\bf k})-
i\sigma h_y({\bf k})\bigl)\partial_x
\rho_{-\sigma\sigma{\prime}}({\bf R},{\bf k},t)
+\partial_{k_x}\bigl(h_x({\bf k})+i\sigma^{\prime}h_y({\bf k})\bigr)
\partial_x\rho_{\sigma-\sigma^{\prime}}
({\bf R},{\bf k},t)\Bigr]\nonumber\\
&&+{1\over 2}\sum_{\sigma_1}\Bigl[
\partial_{k_x}\Sigma_{\sigma\sigma_1}({\bf R},{\bf k},t)
\partial_x\rho_{\sigma_1\sigma^{\prime}}({\bf R},{\bf k},t)
+\partial_x\rho_{\sigma\sigma_1}({\bf R},{\bf k},t)
\partial_{k_x}\Sigma_{\sigma_1\sigma^{\prime}}({\bf R},{\bf k},t)
\Bigr].
\end{eqnarray}
\narrowtext
\noindent It is therefore noted that  
the corresponding coefficients of $\partial_x\rho_{\sigma\sigma^{\prime}}$, 
$\partial_x\rho_{-\sigma\sigma^{\prime}}$ and
$\partial_x\rho_{\sigma-\sigma^{\prime}}$ in the Bloch eqautions  are
\begin{eqnarray}
  \label{eq:cfr}
  && {k_x\over m^{\ast}}+{1\over
    2}\partial_{k_x}[\Sigma_{\sigma\sigma}({\bf R}, {\bf
    k},t)+\Sigma_{\sigma^{\prime}\sigma^{\prime}}({\bf R},{\bf k},t)]\;,\\
  &&{1\over 2}\partial_{k_x}\bigl\{[h_x({\bf k})-i\sigma h_y({\bf
    k})]/2+\Sigma_{\sigma-\sigma}({\bf R}, {\bf k},t)\big\}\;, 
  \\ 
&&{1\over 2}\partial_{k_x}\bigl\{[h_x({\bf k})+i\sigma^{\prime} h_y({\bf k})]/2
  +\Sigma_{-\sigma^{\prime}\sigma^{\prime}}({\bf R},{\bf k},t)\bigr\}\;,
  \label{eq:crr}
\end{eqnarray}
respectively.  They are all $k$-dependent.  
Hence, similar to the simplified model, the
interference effect is also important in the full kinetic equation.
The kinetic equations (\ref{eq1}) and the Poisson equation
(\ref{poission}), together with 
the boundary conditions (\ref{eq:fk_0}) and (\ref{eq:rho_0}) can be
solved numerically in an iterative manner to achieve the stationary
solution.\cite{floyd_jap_1994,coz_thesis} The numerical results for
a typical QW with width $a=7.5$\ nm, boundary spin polarization
$N_{1/2}(0)=2.05\times 10^{11}$\ cm$^{-2}$ and
$N_{-1/2}(0)=1.95\times 10^{11}$\ cm$^{-2}$ at temperature $T=200$K
 are  plotted in Fig.\ 2. In this computation, 
we only take into account the scattering due to longitude optical 
(LO) phonon. It  can be seen from the figure
that the surplus of the spin up electrons decreases rapidly along
the diffusion direction, similar to the simplified
model shown above. 
\vskip -0.3cm
\begin{figure}[htb]
  \psfig{figure=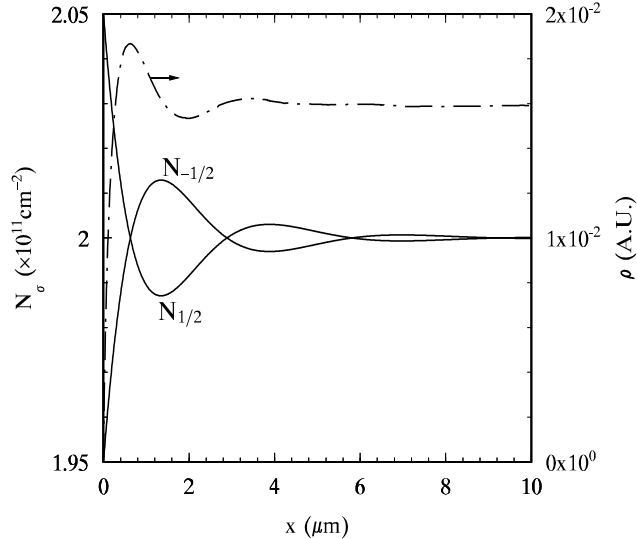,width=9.1cm,height=8.cm,angle=0}
  \caption{Electron densities of up spin and down spin (solid curves)
    and incoherently summed spin coherence $\rho$ (dashed curve) versus
    the diffusion length $x$. $B=1$\ T.  Note the scale of the spin
    coherence is on 
    the right side of the figure.}
\end{figure}
\vskip -0.3cm
\begin{figure}[htb]
\psfig{figure=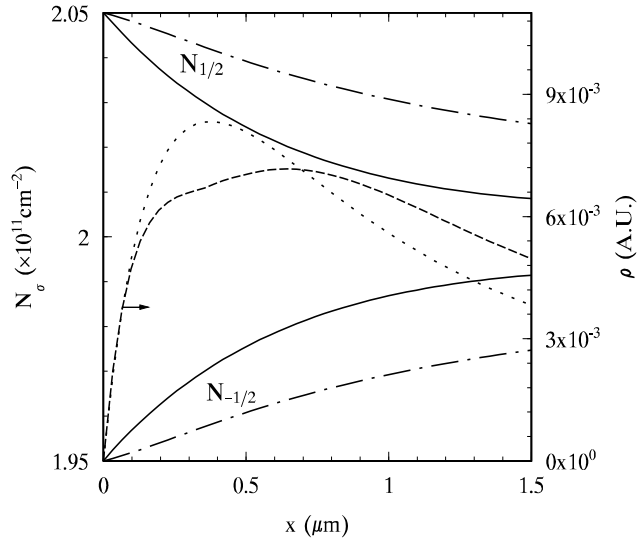,width=9.1cm,height=8.cm,angle=0}
\caption{Electron densities of up spin and down spin and 
 the incoherently summed spin coherence versus the diffusion length $x$.  
Solid curves and dashed curve: $N_\sigma$ and $\rho$ from the full Bloch equations;
Dash-dotted curves and  dotted curve: $N_\sigma$
and $\rho$ from the 
equations without the interference effect. Note the scale of the spin
coherence is on the right side of the figure.}
\end{figure}

The fast decay above is understood mainly generated by the 
decoherence from the interference effect due to the inhomogeneous  broadening.
Other dephasing effects such as those caused by the DP terms in
 Eqs.\ (\ref{eq:fc}) and (\ref{eq:rhoc}) as well as  the spin conserving
LO phonon  scattering also contribute to the decay. Besides, we pointed out that
the inhomogeneous broadening effect combined with spin-conserving
scattering can also cause spin dephasing.\cite{wu2} 
 Therefore, the above mentioned 
inhomogeneous broadening may also cause spin dephasing in
the presence of LO phonon scattering.  To compare the decoherence due
to interference and the dephasing due to the DP term 
together with the scattering, we remove the interference effect in
the transport  equations by replacing $k$ in the coefficients [Eqs.\
(\ref{eq:cfr})-(\ref{eq:crr})] with $k=k_F$.  Here $k_F$ represents
the Fermi wavevector.  Therefore, if there is any decay of spin polarization
along the diffusion direction, it  comes from the spin dephasing.
The numerical result is  plotted in Fig.\ 2. It is shown clearly that the
decay of spin polarization due to the dephasing effect alone
(dash-dotted curves) is much slower than that due to the decoherence
(interference) effect. In the figure we also plot the corresponding
incoherently summed spin coherences $\rho$. One can see from the
figure that both coherences $\rho$ decay slowly and their decay rates
are comparable when $x>1$\ $\mu$m. 
This further justifies what mentioned above that the fast decay of
the spin polarization is mainly due to the interference effect.

In conclusion, we have set up many-body kinetic Bloch equations with
spacial inhomogeneity. We reexamined the wildly adopted
quasi-independent electron model and pointed out an important
many-body spin decoherence effect which is missing in  the 
single electron model. The new decoherence effect  is based 
on an interference effect
along the diffusion direction in spin transport due to the so 
called inhomogeneous broadening effect. We have shown that this
inhomogeneous broadening effect 
 can alone cause spin decoherence, 
even without the scattering and that the resulting decoherence
is more important than the dephasing effect due to  both the DP term 
and the scattering.  
Our study shows the inadequacy of the quasi-independent electron
model. Therefore, it is important to use the full many-body theory to
study the spin transport.

MWW is supported by the ``100 Person Project" of Chinese Academy of
Sciences and Natural Science Foundation of China under Grant
No. 10247002. He is also partially supported by Cooperative Excitation
Project of ERATO (JST).

\end {multicols}
\end {document}